# Pressure-induced magnetic properties of quasi-2D $Cr_2Si_2Te_6$ and $Mn_3Si_2Te_6$

Rubyann Olmos[1†], Po-Hao Chang[1], Prakash Mishra[2], Rajendra R. Zope[1], Tunna Baruah[1], Cedomir Petrovic[3], Yu Liu[3‡], Srinivasa R. Singamaneni[1*]

[1]Department of Physics, The University of Texas at El Paso, El Paso, TX 79968, USA

[2]Computational Science Program, The University of Texas at El Paso, El Paso, TX 79968, USA

[3]Condensed Matter Physics and Materials Science Department, Brookhaven National Laboratory, Upton, New York 11973, USA

Recently, the pressure has been used as external stimuli to induce structural and magnetic phase transitions in many layered quantum materials whose layers are linked by van der Waals forces. Such materials with weakly held layers allow relatively easy manipulation of the superexchange mechanism and lead to novel magnetic behavior. Using the hydrostatic pressure as a disorderless means to manipulate the interlayer coupling, we applied pressure on two quasi-2D sister compounds, namely, $Cr_2Si_2Te_6$ (CST) and $Mn_3Si_2Te_6$ (MST), up to ~1 GPa. Magnetic property measurements with the application of pressure revealed that the ferromagnetic transition temperature decreases in CST while the opposite trend occurs for the ferrimagnet MST. In MST, the magnetization decreases with the increase in the pressure, and such trend is not clearly noticed in CST, within the pressure range studied. Theoretical calculations showed the overall pressure effect on layer separation, bond angle, and exchange coupling, strongly influencing the change in subsequent magnetic characteristics. Exchange coupling in $Mn_3Si_2Te_6$ is strongly frustrated and the first nearest neighbor interaction is the most dominant of the components with the strongest pressure dependence. Whereas, in $Cr_2Si_2Te_6$, the exchange coupling parameters exhibit very little dependence on the pressure. This combined experimental and theoretical work has the potential to expand to other relevant quantum materials.

*Corresponding author: srao@utep.edu

[†]Present address: Department of Physics and Astronomy, Rice University, Houston, TX 77005, USA.

[‡]Present address: Los Alamos National Laboratory, Los Alamos, NM 87545, USA.

## I. INTRODUCTION

Since the magnetic properties of bulk (quasi-2D) crystals serve as a basis for the understanding of magnetic phenomenon in reduced dimensions (true 2D limit), a profound knowledge and understanding of these compounds is evidently needed at the quasi-2D level, especially considering that magnetic mono- and bi-layers of a wide range of 2D quantum magnets have now become accessible.[1–5] Therefore, expanding the knowledge on the magnetic behavior in quasi-2D materials has the great potential in driving the boundaries of materials towards future use in spin electronic based devices as well as in studying the exceptional quantum properties. The weak interlayer-bonding present in vdW crystals and their extreme sensitivity toward the external perturbation have allowed the scientists to explore their interesting physical properties.

In the recent past, external stimuli have been employed to tune the magnetic states in (quasi) 2D materials via photoexcitation, proton irradiation, strain, gating, doping, pressure, and intercalation.[5–15] Among all, hydrostatic pressure has been used as a disorder-less procedure to tune the physical properties of vdW compounds. Pressure on a system of weakly coupled layers affects bonding and interlayer coupling without destroying the crystal.[2] For instance, a remarkable tuning of the magnetic properties has been seen in $Cr_2Ge_2Te_6$, where approximately 9 GPa of pressure significantly increased the Curie temperature ($T_C$) to a value greater than 250 K, a jump of nearly 184 K from its ambient conditions.[16] Previous high-pressure ($P$ > 1 GPa to 47 GPa) studies conducted on $CrSiTe_3$ reported structural transformations, metallization, spin reorientation transition, and a ferromagnetic to paramagnetic transition with the emergence of superconductivity.[17,18] The most recent work on few-layered $CrSiTe_3$ reported a dramatic increase in the $T_C$ by about 100 K with 7.8 GPa of pressure application.[19] Moreover, it's sister compound, $Mn_3Si_2Te_6$, recently gained a great deal of interest due to the observance of exceptional colossal magneto resistance.[20–22] More recently, $Mn_3Si_2Te_6$ has shown a pressure-dependence of the colossal magneto resistance, a semiconductor to metal transition, and structural phase transition.[23]

As outlined above, while most of the prior work reported on $Cr_2Si_2Te_6$ (CST) and $Mn_3Si_2Te_6$ (MST) focused on the high-pressure properties, the knowledge on their magnetic properties, particularly, at lower (< 1 GPa) pressure is limited. In this work, we examined the influence of pressure on the magnetic properties of two bulk van der Waals materials, CST and MST. The application of pressure revealed that the ferromagnetic transition temperature decreases in CST while the opposite trend occurs for the ferrimagnet, MST. In MST, the magnetization decreases with the increase in the pressure, and such trend is not clearly noticed in CST, within the pressure range studied. Density functional theory (DFT) calculations give further insights into the pressure dependence of exchange coupling and magnetocrystalline anisotropy energy in these materials.

The semiconducting material, CST, exhibits a $T_C$ of 32 K with an easy-axis along the *c*-axis.[24] CST crystallizes with rhombohedral symmetry in a honeycomb network with space group R-3. The Cr ions are arranged in hexagonal planes stacking along the *c*-axis with one-third of layers being composed of $Si_2Te_6$. Each $Cr^{3+}$ atom ($S$ = 3/2) is octahedrally coordinated with Te that are edge sharing. The insulating compound, MST, is a ferrimagnet whose antiferromagnetic contribution stems from frustration between the three nearest neighbor interactions between Mn1 (multiplicity of two) and Mn2 sites. MST is often referred to as the sister compound to $Cr_2Si_2Te_6$ as layers are composed of $MnTe_6$ octahedra sharing an edge with the *ab* plane at the Mn1 site and with Si-Si dimers. However, MST differs in that one-third of the Mn atoms link the layers together by filling the octahedral holes at the Mn2 site.[25] MST has a trigonal crystal structure with space group no. 163 [26,27] with $T_C$ between 73 to 78 K.[10,25,28]

## II. METHODS

**Experimental Details**

Bulk CST and MST crystals were synthesized using a self-flux technique as previously outlined.[28,29] Magnetic measurements were performed using a Quantum Design MPMS 3 with Superconducting Quantum Interference Device (SQUID) magnetometer. Isothermal magnetization measurements were taken at 2 K with a ± 7 T magnetic field. Zero-field cool (ZFC) temperature dependent magnetization was performed from 2–150 K with a measuring field of 500 Oe. Hydrostatic pressure was applied using a BeCu Quantum Design piston cell. The pressure transmitting medium is Daphne (Silicon) oil and a Pb manometer was used to monitor the pressure in the cell. Compression of the cell length was increasingly applied and only depressurized after the data collection was completed. For CST, we perform pressure measurements with the magnetic field applied parallel to the *c*-axis (H ∥ *c*), out-of-plane measurements. For MST, we fixed the sample along its magnetic easy axis, H ∥ *ab*. For additional information on mounting procedures and how pressure was determined see the supplementary information.

**Computational Details**

The structural parameters were optimized using Vienna *ab* initio Simulation Package (VASP) [30,31] within projector augmented wave (PAW) method.[32,33] The Perdew-Burke-Enzerhof (PBE) [34] generalized gradient approximation was employed to describe exchange correlation effects. To better account for the interlayer van der Waals (vdW) forces, optB88-vdW [35,36] form of the non-local vdW functional was used. The cutoff energy of 500 eV was chosen for the plane wave basis set.[37] The structure optimizations were performed until the forces on each atom were less than 5 x $10^{-3}$ meV/A. The hydrostatic pressure effect is included using the PSTRESS tag, which adds the stress to the stress tensor and an energy generated from the external pressure.

Once the optimized structures were obtained, we use the OpenMX code [38] for calculations of magnetic properties. In these calculations, core electrons are replaced with norm-conserving pseudopotential [39] with energy cutoff 300 Ry. To improve the description for localized *d*-electrons in Mn and Cr, fully localized limit DFT+$U$ is used.[40,41] The $U$ parameters for CST and MST are taken from literature.[19,25] Green function method implemented in OpenMX 3.9 [42] was used to calculate exchange coupling constants $J_i$ for up to the 3rd nearest neighbor (NN). In this approach, the exchange coupling between any pair of magnetic sites can be directly calculated from a single magnetic state. For MAE, we first perform a full self-consistent calculation without spin-orbit coupling (SOC) to obtain the charge and spin densities. The SOC is subsequently included perturbatively for the configurations with the magnetization aligned in-plane and out-of-plane. The MAE was then determined as the difference in the total band energy between the two.[43]

## III. RESULTS AND DISCUSSION

First, we present and discuss our experimental data gathered on CST, where we applied the pressure up to 0.686 GPa. Isothermal magnetization (M-H) measurements were conducted along the easy axis (H ∥ *c*). Figure 1(a) plots the M-H data collected on CST at 2 K with ± 7 T with various pressures in the range between 0 to 0.686 GPa. A small decrease (34.16 to 34.11 emu/g) in the saturation magnetization ($M_S$) measured at the temperature of 2 K, and the magnetic field of 7 T (maximum limit of our magnetometer) is noted as the pressure is increasingly applied. The zero-field cooled (ZFC) temperature variation of magnetization (M-T) collected at various pressures is plotted in Figure 1(b) with a 500 Oe magnetic field. This plot shows a similar decrease in the magnetization in the ferromagnetic phase (below 34 K). Most notably, we observed that M-T curves shifted to the left, indicating a decrease in the magnetic phase transition temperature, $T_C$, as pressure is increased. Figure 1(c) displays the derivative of the magnetization with respect to the temperature. It is observed at the maximum applied pressure (0.686 GPa) that $T_C$ decreases to 28 K, in addition to a drastic decrease in the magnitude of the minimum(s) at the largest pressure. This shrinking of the derivative curve, along with the suppression of $T_C$ and magnetization indicate a decrease in the overall magnetic ordering as similarly seen in pressure study involving $Fe_3GeTe_2$.[44] Figure 1(d) displays the magnetization taken at 7 T and $T_C$ as a function of pressure.

Turning our attention to the experiment on MST, we applied hydrostatic pressure up to 0.90 GPa with magnetic field applied along the easy axis (H ∥ *ab*). Figure 2(a) displays the isothermal magnetization at 2 K and at ± 7 T, collected as a function of applied pressure. At first glance, the $M_s$ at 7 T follows a decreasing trend as pressure is increased to 0.90 GPa. Figure 2(b) shows the temperature (2-150 K) variation of ZFC magnetization (measuring field: 500 Oe) at various applied pressures (0 to 0.9 GPa). Interestingly, we noticed an apparent shift of the M-T curves towards the right as the pressure is steadily increased. The $T_C$ is estimated in Figure 2(c) through the magnetization derivative curves, which shows an increase up to 84.52 K at 0.9 GPa, an 11 K jump from the pristine state. Although the relationship between magnetization and $T_C$ as a function of pressure is not very clear with the current experimental results, see Figure 2(d), this increasing trend in $T_C$ has recently been observed for even higher pressures in MST with results showing a remarkable jump in $T_C$ up to ~210 K at 8.1 GPa.[23]

To gain additional insight, we also performed first-principle based calculations for both CST and MST to study the influence of pressure on their magnetic properties. Earlier theoretical calculations have established that the ferromagnetic phase is the lowest energy spin structure for CST. Our calculations are also carried out on the FM phase. Figure 3 shows the in-plane exchange coupling parameters $J_i$ (*i* = 1,2,3) and out-of-plane parameter $J_{out}$ that are calculated in this work as a function of pressure for CST as shown in Figure 4(a). Due to the nature of our approach where $J_i$'s are calculated for every given pair of sites independently, we considered a few nearest neighbors. Fig. 4(a), $J_1$, $J_2$, and $J_3$ are the exchange couplings between the first, second, and the third nearest neighbors, respectively corresponding to Cr-Cr distances of 3.95, 6.85 and 7.91 Å. The calculated exchange coupling parameters as a function of pressure are shown in Fig. 4(a) for CST. These parameters are calculated both at the GGA and GGA+U level. The nearest neighbor in-plane coupling $J_1$ is the most dominant interaction followed by $J_3$. All the calculated exchange couplings including the interlayer coupling are ferromagnetic in nature. The in-plane exchange coupling is dominated by $J_1$ and considering that there exist three pairs of nearest neighbors within a unit cell, the exchange couplings strongly favor a ferromagnetic state. Fig. 4(a) shows that the couplings remain ferromagnetic within the pressure range considered here in accord with earlier results of Zhang et al. [19]. The coupling parameters do not change significantly under the application of pressure unto 1 GPa. Pressure reduces the distances between the Cr ions and as a result the FM coupling is enhanced but this change is small within the range of pressure used here.

It has been shown that in CST different layers are only weakly coupled.[19,45] We have calculated the coupling between the interlayer nearest neighbors which are at a distance of 6.86 Å. Our results also show that the layers are ferromagnetically coupled but the coupling is weak. Weak interlayer coupling has also been found in the earlier study [45] where an effective Heisenberg spin Hamiltonian with up to 3$^{rd}$ NN interlayer exchange coupling was constructed. Within this pressure range (0 to 1 GPa), the interlayer ($J_i$) and intralayer ($J_{out}$) couplings exhibit very little dependence on the pressure. We also observed that the empirical Hubbard $U$ value does not change the qualitative behavior in pressure dependence. Earlier theoretical studies of Zhang et al. [19] have shown that the magneto-crystalline anisotropy energy slightly decreases with pressure in the range 0-8 GPa. Our results also show (Fig. 4(b)) that the magneto crystalline anisotropy energy decreases slowly with pressure in the range 0-1 GPa. The difference between the total band energies with in-plane and out-of-plane magnetization directions showed that the system has an out-of-plane easy axis of anisotropy.

Due to the triangular arrangement in MST as shown in Fig. 5, the system is rather frustrated and several magnetic ground states are close in energy.[25] To verify that, we also consider the total energy differences of several magnetic states: two ferrimagnetic states (FI1 and FI2), two anti-ferromagnetic (AF1 and AF2), and a ferromagnetic state (FM). The ground state is found to be ferrimagnetic (FI1) with the spins of Mn1 and Mn2 sites oppositely aligned as shown in Fig. 5. In the FI2 state, the Mn1 ions have spins antiferromagnetically aligned, while half of the Mn1-Mn2 spins are antiferromagnetically aligned, and the other half are ferromagnetically aligned. In AF1 and AF2 the Mn1 spins are oppositely aligned. In AF1 the Mn2 spin is aligned similar to the nearest Mn1 while in AF2 the Mn2 spins are oppositely aligned to the nearest Mn1.[25] Table I shows the relative energies of the magnetic states calculated with GGA and GGA+$U$ compared to the ferrimagnetic ground state (FI1) which are in agreement with previous results obtained using the linearized augmented plane-wave code.[25] These calculations established that the FI1 is the ground state. Further calculations with hydrostatic pressure were done on the FI1 state only.

The three exchange coupling parameters calculated here are depicted in Fig. 5. Here $J_1$, $J_2$, and $J_3$ are the couplings between nearest neighbors Mn1-Mn2, Mn1-Mn1, and second nearest neighbors Mn1-Mn2. Figure 6(b) shows the pressure dependence of exchange couplings calculated with for both $U = 0$ and $U = 3.0$ eV. Use of the $U$ reduces the strengths of the exchange couplings in MST but the overall qualitative behavior remains same. Similar to the previous work [25], for both $U$ values, all the three components favor AF, which again confirms the magnetically frustrated nature of the system due to the competition between neighboring pairs. Application of pressure affects mostly the $J_1$ parameter which is the most dominant interaction that favors the moments of the corresponding Mn1-Mn2 pairs to be antiferromagnetically ordered. As pressure increases the orbitals become more delocalized which will favor antiferromagnetic interactions. On the other hand, the $J_2$ and $J_3$ are similar in strength at the GGA level and both favors antiferromagnetic ordering of the respective pairs leading to a frustrated system. With +$U$, the $J_3$ is larger than $J_2$ which stabilizes the FI1 state. The pressure dependence of $J_2$ and $J_3$ is very small possibly due to the frustration in the system. Increase of $J_1$ with pressure indicates that the FI1 state is stabilized. Fig. 6(b) we show the magnetic anisotropy energy as a function of applied pressure. Our calculations predict an easy-plane anisotropy with MAE around 0.68 meV/Mn at zero pressure which is in close agreement with previous work Ref. [25] and the magnitude decreases monotonically by about 15% at P = 1.0 GPa.

To qualitatively understand the trend of Curie temperature evolution with pressure, we employ the following mean-field approximation method [46]:

$$T_C = \frac{2}{3} \frac{\Delta E}{k_B N}$$

where $\Delta E$ is the total energy difference between the ordered and the disordered local moment states, $N$ is the total number of spin centers in the supercell, and $k_B$ is Boltzmann's constant. In our approach, we approximate the $\Delta E$ as the total energy difference between the ground and the low-lying antiferromagnetic state with zero net spin moment. As shown in Table 1, the relative ordering between the AF1 and AF2 changes with GGA+$U$ compared to GGA. Relative energies of both the AF1 and AF2 increases compared to the ground state which indicates that the $T_C$ will also increase with pressure in accord with the experimental results. The stabilization of the ground state as indicated by the increase in $J_1$ also indicates that the energy difference between the ground and the paramagnetic state is likely to increase leading to rise in $T_C$.

### IV. CONCLUSIONS

In this work, we investigated the pressure-dependent magnetic properties of two quasi-2D magnets, namely, $Cr_2Si_2Te_6$ and $Mn_3Si_2Te_6$ by employing the Quantum Design Magnetometer coupled with theoretical calculations. In the case of CST, we find that the Curie temperature is decreased from 34 K to 28 K as the pressure is increased from 0 GPa to 0.686 GPa, though the saturation magnetization is marginally decreased within the pressure range applied. Moreover, it will be important to observe the coercivity and the interplanar distance to corroborate the results from previous high-pressure studies. It also seems that the literature will benefit from a comprehensive magnetization study in the high-pressure regime to bring together the results of pressure enhanced ferromagnetism in Ref. [19] up to 7.8 GPa and a structural transition causing a ferromagnetic to paramagnetic transition leading to superconductivity at 7.5 GPa,[18] to a spin reorientation at ~ 6 GPa inducing metallization [17]. Quite intriguingly, in the case of MST, the $T_C$ is found to increase by 11 K upon increase of pressure from 0 to 0.9 GPa. Recent rich phenomena has been demonstrated in high pressure studies on MST such as pressure-induced semiconductor to metal transition between 1.5 and 2.5 GPa and using pressure to control the apparent colossal magnetoresistance that is present in this material.[20,23] DFT calculations at GGA+U level indicates that the nearest neighbor exchange couplings and MAE change monotonically with pressure up to 1 GPa. A qualitative estimate of the change in $T_C$ with pressure shows that $T_C$ is likely to increase with pressure in accord with experimental results. To gain deeper insights and further expand this work, we are currently performing the synchrotron-based high-pressure X-ray magnetic circular dichroism measurements and pressure-dependent Raman measurements.

**ACKNOWLEDGMENTS**

This material is based upon work supported by the National Science Foundation Graduate Research Fellowship Program under Grant No. 1842494. Any opinions, findings, and conclusions or recommendations expressed in this material are those of the author(s) and do not necessarily reflect the views of the National Science Foundation. S.R.S., R.O., P.M., and T.B. acknowledge support from the NSF-DMR (Award No. 2105109). SRS acknowledges support from NSF-MRI (Award No. 2018067). PH.C., T.B., and R.R.Z. acknowledge support from the US Department of Energy, Office of Science, Office of Basic Energy Sciences, as part of the Computational Chemical Sciences Program under Award No. DE-SC0018331. Support for computational time at the Texas Advanced Computing Center directly and through NSF Grant No. TG-DMR090071 and NERSC is gratefully acknowledged. Work at Brookhaven National Laboratory is supported by the U.S. DOE under Contract No. DESC0012704 (materials synthesis).

**REFERENCES**

1. Duong DL, Yun SJ, Lee YH. van der Waals Layered Materials: Opportunities and Challenges. *ACS Nano* (2017) **11**:11803–11830. doi:10.1021/acsnano.7b07436

2. Ajayan P, Kim P, Banerjee K. Two-dimensional van der Waals materials. *Physics Today* (2016) **69**:38–44. doi:10.1063/PT.3.3297

3. Gong C, Zhang X. Two-dimensional magnetic crystals and emergent heterostructure devices. *Science* (2019) **363**:eaav4450. doi:10.1126/science.aav4450

4. Huang B, Clark G, Navarro-Moratalla E, Klein DR, Cheng R, Seyler KL, Zhong D, Schmidgall E, McGuire MA, Cobden DH, et al. Layer-dependent ferromagnetism in a van der Waals crystal down to the monolayer limit. *Nature* (2017) **546**:270–273. doi:10.1038/nature22391

5. Wang QH, Bedoya-Pinto A, Blei M, Dismukes AH, Hamo A, Jenkins S, Koperski M, Liu Y, Sun Q-C, Telford EJ, et al. The Magnetic Genome of Two-Dimensional van der Waals Materials. *ACS Nano* (2022) **16**:6960–7079. doi:10.1021/acsnano.1c09150

6. Padmanabhan P, Buessen FL, Tutchton R, Kwock KWC, Gilinsky S, Lee MC, McGuire MA, Singamaneni SR, Yarotski DA, Paramekanti A, et al. Coherent helicity-dependent spin-phonon oscillations in the ferromagnetic van der Waals crystal $CrI_3$. *Nat Commun* (2022) **13**:4473. doi:10.1038/s41467-022-31786-3

7. Singamaneni SR, Martinez LM, Niklas J, Poluektov OG, Yadav R, Pizzochero M, Yazyev OV, McGuire MA. Light induced electron spin resonance properties of van der Waals $CrX_3$ (X = Cl, I) crystals. *Applied Physics Letters* (2020) **117**:082406. doi:10.1063/5.0010888

8. Olmos R, Alam S, Chang P-H, Gandha K, Nlebedim IC, Cole A, Tafti F, Zope RR, Singamaneni SR. Pressure dependent magnetic properties on bulk $CrBr_3$ single crystals. *Journal of Alloys and Compounds* (2022) **911**:165034. doi:10.1016/j.jallcom.2022.165034

9. Martinez LM, Iturriaga H, Olmos R, Shao L, Liu Y, Mai TT, Petrovic C, Walker ARH, Singamaneni SR. Enhanced magnetization in proton irradiated Mn3Si2Te6 van der Waals crystals. *Applied Physics Letters* (2020) **116**:172404. doi:10.1063/5.0002168

10. R. Olmos, J. Delgado, H. Iturriaga, L. M. Martinez, C. L. Saiz, L. Shao, Y. Liu, C. Petrovic, S. R. Singamaneni. Critical phenomena of the layered ferrimagnet Mn3Si2Te6 following proton irradiation. *J Appl Phys 130, 013902 (2021)* doi:https://doi.org/10.1063/5.0056387

11. Mak KF, Shan J, Ralph DC. Probing and controlling magnetic states in 2D layered magnetic materials. *Nat Rev Phys* (2019) **1**:646–661. doi:10.1038/s42254-019-0110-y

12. Jiang S, Li L, Wang Z, Mak KF, Shan J. Controlling magnetism in 2D CrI3 by electrostatic doping. *Nature Nanotech* (2018) **13**:549–553. doi:10.1038/s41565-018-0135-x

13. Wang Z, Zhang T, Ding M, Dong B, Li Y, Chen M, Li X, Huang J, Wang H, Zhao X, et al. Electric-field control of magnetism in a few-layered van der Waals ferromagnetic semiconductor. *Nature Nanotech* (2018) **13**:554–559. doi:10.1038/s41565-018-0186-z

14. Wang N, Tang H, Shi M, Zhang H, Zhuo W, Liu D, Meng F, Ma L, Ying J, Zou L, et al. Transition from Ferromagnetic Semiconductor to Ferromagnetic Metal with Enhanced Curie Temperature in Cr2Ge2Te6 via Organic Ion Intercalation. *J Am Chem Soc* (2019) **141**:17166–17173. doi:10.1021/jacs.9b06929

15. Liu Y, Susilo RA, Lee Y, Abeykoon AMM, Tong X, Hu Z, Stavitski E, Attenkofer K, Ke L, Chen B, et al. Short-Range Crystalline Order-Tuned Conductivity in Cr2Si2Te6 van der Waals Magnetic Crystals. *ACS Nano* (2022) **16**:13134–13143. doi:10.1021/acsnano.2c06080

16. Bhoi D, Gouchi J, Hiraoka N, Zhang Y, Ogita N, Hasegawa T, Kitagawa K, Takagi H, Kim KH, Uwatoko Y. Nearly Room-Temperature Ferromagnetism in a Pressure-Induced Correlated Metallic State of the van der Waals Insulator CrGeTe 3. *Phys Rev Lett* (2021) **127**:217203. doi:10.1103/PhysRevLett.127.217203

17. Xu K, Yu Z, Xia W, Xu M, Mai X, Wang L, Guo Y, Miao X, Xu M. Unique 2D–3D Structure Transformations in Trichalcogenide CrSiTe 3 under High Pressure. *J Phys Chem C* (2020) **124**:15600–15606. doi:10.1021/acs.jpcc.0c03931

18. Cai W, Sun H, Xia W, Wu C, Liu Y, Liu H, Gong Y, Yao D-X, Guo Y, Wang M. Pressure-induced superconductivity and structural transition in ferromagnetic CrSiTe 3. *Phys Rev B* (2020) **102**:144525. doi:10.1103/PhysRevB.102.144525

19. Zhang C, Gu Y, Wang L, Huang L-L, Fu Y, Liu C, Wang S, Su H, Mei J-W, Zou X, et al. Pressure-Enhanced Ferromagnetism in Layered CrSiTe 3 Flakes. *Nano Lett* (2021) **21**:7946–7952. doi:10.1021/acs.nanolett.1c01994

20. Ni Y, Zhao H, Zhang Y, Hu B, Kimchi I, Cao G. Colossal magnetoresistance via avoiding fully polarized magnetization in the ferrimagnetic insulator Mn3Si2Te6. *PHYSICAL REVIEW B* (2021)6.

21. Zhang Y, Ni Y, Zhao H, Hakani S, Ye F, DeLong L, Kimchi I, Cao G. Control of chiral orbital currents in a colossal magnetoresistance material. *Nature* (2022) **611**:467–472. doi:10.1038/s41586-022-05262-3

22. Liu Y, Hu Z, Abeykoon M, Stavitski E, Attenkofer K, Bauer ED, Petrovic C. Polaronic transport and thermoelectricity in Mn3Si2Te6 single crystals. *PHYSICAL REVIEW B* (2021) **103**:7.

23. Wang J, Wang S, He X, Zhou Y, An C, Zhang M, Zhou Y, Han Y, Chen X, Zhou J, et al. Pressure engineering of colossal magnetoresistance in the ferrimagnetic nodal-line semiconductor Mn 3 Si 2 Te 6. *Phys Rev B* (2022) **106**:045106. doi:10.1103/PhysRevB.106.045106

24. Casto LD, Clune AJ, Yokosuk MO, Musfeldt JL, Williams TJ, Zhuang HL, Lin M-W, Xiao K, Hennig RG, Sales BC, et al. Strong spin-lattice coupling in CrSiTe $_3$. *APL Materials* (2015) **3**:041515. doi:10.1063/1.4914134

25. May AF, Liu Y, Calder S, Parker DS, Pandey T, Cakmak E, Cao H, Yan J, McGuire MA. Magnetic order and interactions in ferrimagnetic Mn3Si2Te6. *PHYSICAL REVIEW B* (2017)10.

26. R. Rimet, C. Schlenker, H. Vincent. A new semiconducting ferrimagnet: A silicon manganese telluride. doi:10.1016/0304-8853(81)90141-4

27. H. Vincent, D. Leroux, D. Bijaoui, R. Rimet, C. Schlenker. Crystal structure of Mn3Si2Te6. doi:10.1016/0022-4596(86)90190-8

28. Liu Y, Petrovic C. Critical behavior and magnetocaloric effect in Mn3Si2Te6. *PHYSICAL REVIEW B* (2018) **98**:6.

29. Liu Y, Petrovic C. Anisotropic magnetic entropy change in Cr 2 X 2 Te 6 ( X = Si and Ge ). *Phys Rev Materials* (2019) **3**:014001. doi:10.1103/PhysRevMaterials.3.014001

30. Kresse G, Furthmüller J. Efficiency of ab-initio total energy calculations for metals and semiconductors using a plane-wave basis set. *Computational Materials Science* (1996) **6**:15–50. doi:10.1016/0927-0256(96)00008-0

31. Kresse G, Furthmüller J. Efficient iterative schemes for *ab initio* total-energy calculations using a plane-wave basis set. *Phys Rev B* (1996) **54**:11169–11186. doi:10.1103/PhysRevB.54.11169

32. Blöchl PE. Projector augmented-wave method. *Phys Rev B* (1994) **50**:17953–17979. doi:10.1103/PhysRevB.50.17953

33. Kresse G, Joubert D. From ultrasoft pseudopotentials to the projector augmented-wave method. *Phys Rev B* (1999) **59**:1758–1775. doi:10.1103/PhysRevB.59.1758

34. Perdew JP, Burke K, Ernzerhof M. Generalized Gradient Approximation Made Simple. *Phys Rev Lett* (1996) **77**:3865–3868. doi:10.1103/PhysRevLett.77.3865

35. Klimeš J, Bowler DR, Michaelides A. Chemical accuracy for the van der Waals density functional. *J Phys: Condens Matter* (2010) **22**:022201. doi:10.1088/0953-8984/22/2/022201

36. Klimeš J, Bowler DR, Michaelides A. Van der Waals density functionals applied to solids. *Phys Rev B* (2011) **83**:195131. doi:10.1103/PhysRevB.83.195131

37. Monkhorst HJ, Pack JD. Special points for Brillouin-zone integrations. *Phys Rev B* (1976) **13**:5188–5192. doi:10.1103/PhysRevB.13.5188

38. Ozaki T. Variationally optimized atomic orbitals for large-scale electronic structures. *Phys Rev B* (2003) **67**:155108. doi:10.1103/PhysRevB.67.155108

39. Vanderbilt D. Soft self-consistent pseudopotentials in a generalized eigenvalue formalism. *Phys Rev B* (1990) **41**:7892–7895. doi:10.1103/PhysRevB.41.7892

40. Anisimov VI, Solovyev IV, Korotin MA, Czyżyk MT, Sawatzky GA. Density-functional theory and NiO photoemission spectra. *Phys Rev B* (1993) **48**:16929–16934. doi:10.1103/PhysRevB.48.16929

41. Ryee S, Han MJ. The effect of double counting, spin density, and Hund interaction in the different DFT+U functionals. *Sci Rep* (2018) **8**:9559. doi:10.1038/s41598-018-27731-4

42. Terasawa A, Matsumoto M, Ozaki T, Gohda Y. Efficient Algorithm Based on Liechtenstein Method for Computing Exchange Coupling Constants Using Localized Basis Set. *J Phys Soc Jpn* (2019) **88**:114706. doi:10.7566/JPSJ.88.114706

43. Chang P-H, Fang W, Ozaki T, Belashchenko KD. Voltage-controlled magnetic anisotropy in antiferromagnetic MgO-capped MnPt films. *Phys Rev Mater* (2021) **5**:054406. doi:10.1103/PhysRevMaterials.5.054406

44. Ding S, Liang Z, Yang J, Yun C, Zhang P, Li Z, Xue M, Liu Z, Tian G, Liu F, et al. Ferromagnetism in two-dimensional $\mathrm{Fe}_{3}\mathrm{Ge}\mathrm{Te}_{2}$; Tunability by hydrostatic pressure. *Phys Rev B* (2021) **103**:094429. doi:10.1103/PhysRevB.103.094429

45. Zhang J, Cai X, Xia W, Liang A, Huang J, Wang C, Yang L, Yuan H, Chen Y, Zhang S, et al. Unveiling Electronic Correlation and the Ferromagnetic Superexchange Mechanism in the van der Waals Crystal CrSiTe 3. *Phys Rev Lett* (2019) **123**:047203. doi:10.1103/PhysRevLett.123.047203

46. Shahjahan M, Toyoda M, Oguchi T. Ferromagnetic Half Metallicity in Doped Chalcopyrite Semiconductors Cu(Al $_{1-x}A_x$ )Se $_2$ ( $A = 3d$ Transition-Metal Atoms). *J Phys Soc Jpn* (2014) **83**:094702. doi:10.7566/JPSJ.83.094702

**FIGURES**

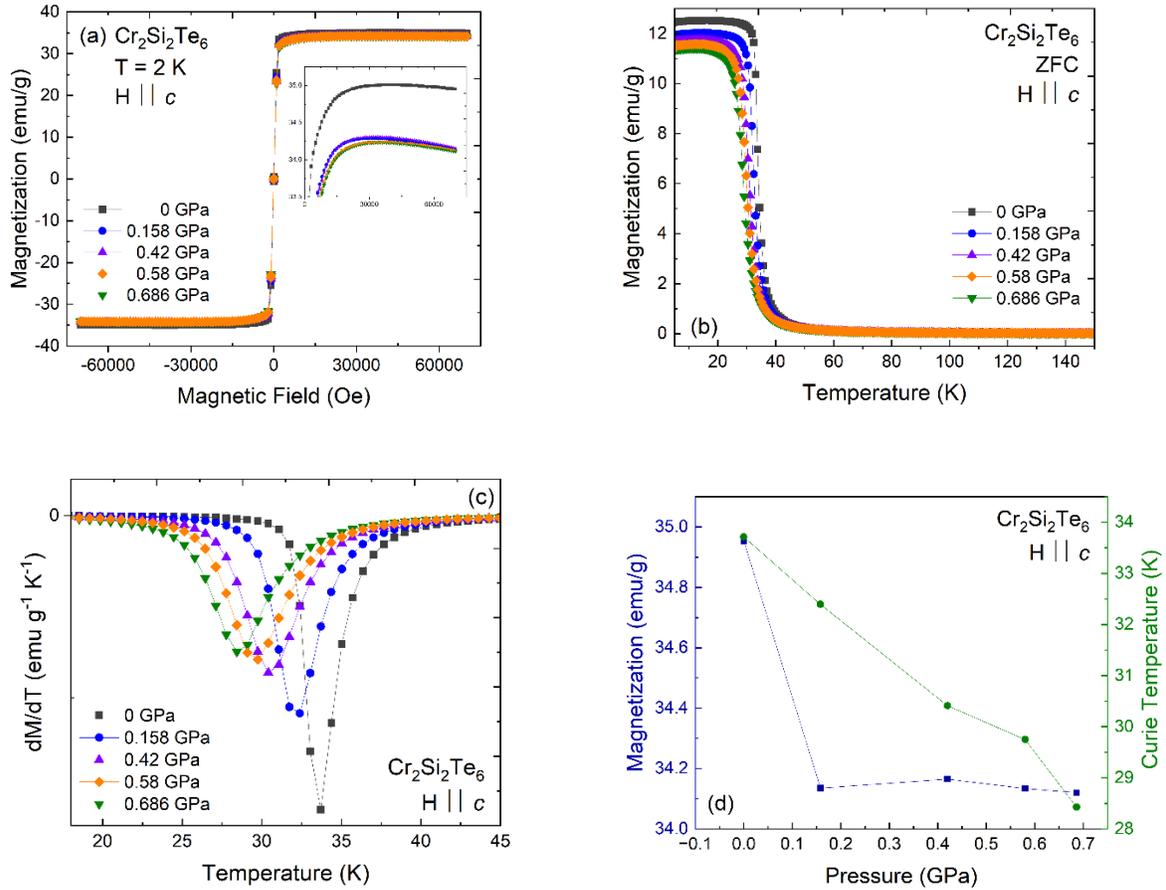

**Figure 1** Experimental data collection on $Cr_2Si_2Te_6$ with application of pressure. **(a)** The isothermal magnetization for the easy axis direction taken at 2 K with ± 7 T. **INSET:** Close up of the magnetization in the 1$^{st}$ quadrant. **(b)** Zero field cool temperature dependent magnetization with a 500 Oe magnetic field in the easy axis (H || c). **(c)** Derivative of the magnetization with respect to temperature shown for all applied pressures. **(d)** Magnetization at 7 T and the ferromagnetic transition temperature plotted as a function of pressure.

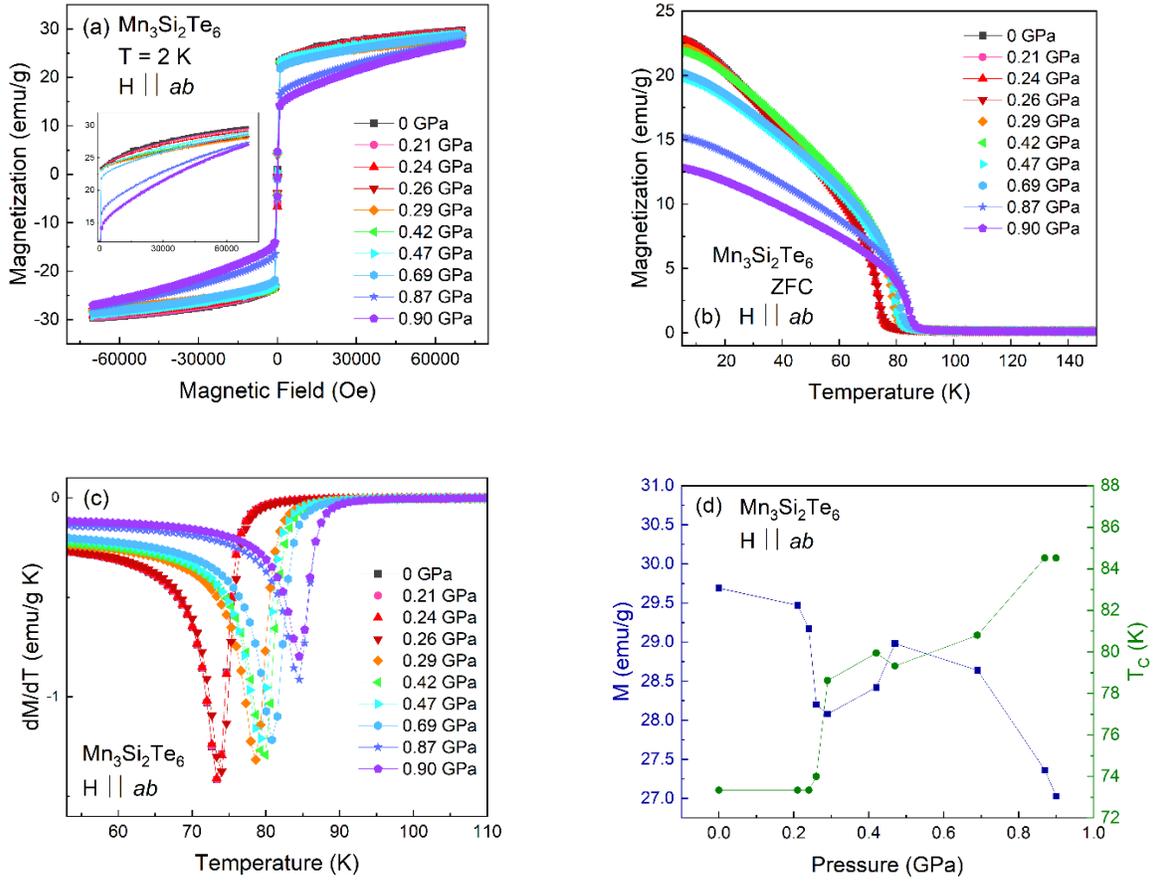

**Figure 2** Experimental data collection on $Mn_3Si_2Te_6$ with application of pressure. **(a)** The isothermal magnetization for the easy axis direction taken at 2 K with ± 7 T. **INSET:** A close up of the magnetization in the 1st quadrant. **(b)** Zero field cool (ZFC) temperature dependent magnetization with a 500 Oe magnetic field in the easy axis (H ∥ ab). **(c)** Derivative of the magnetization with respect to temperature shown for all applied pressures. **(d)** Magnetization at 7 T and $T_C$ plotted as a function of increasing pressure up to 0.9 GPa.

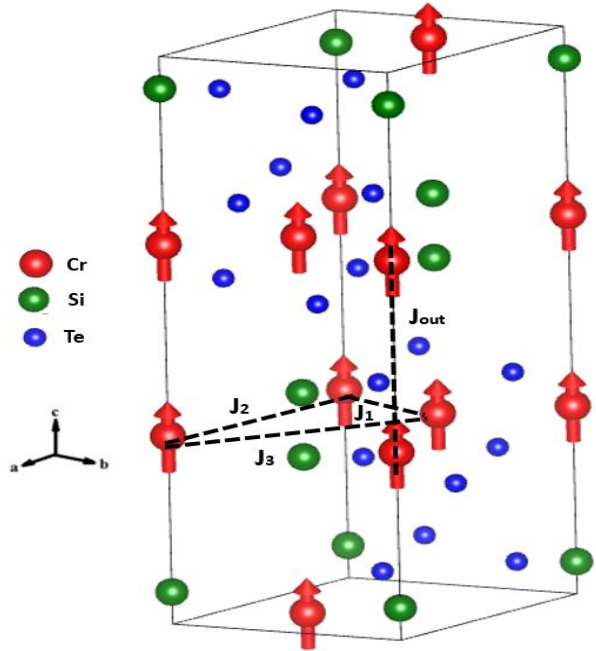

**Fig. 3.** The crystal structure and the spin alignment in CrSiTe$_3$. The calculated *J* parameters between the first, second, and the third in-plane nearest Cr neighbors and the interlayer *J*$_{out}$ are depicted with broken lines.

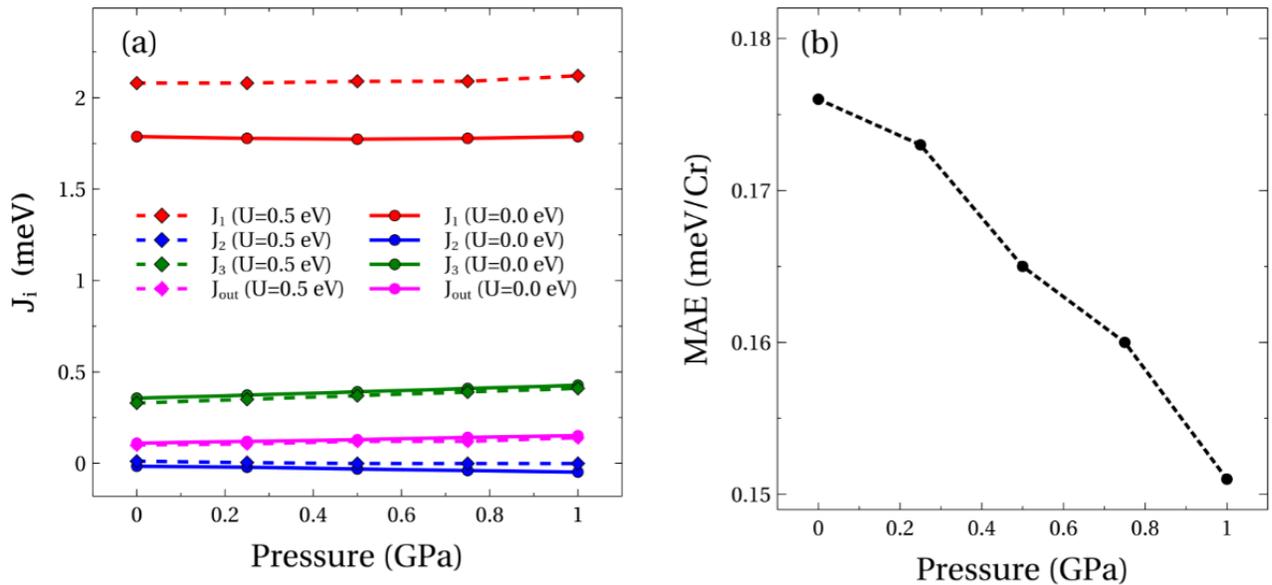

**Figure 4 (a)** Shows the $J_i$ as functions of pressure for CST. Our data suggest that the in-plane exchange coupling is dominated by $J_1$ which strongly favors FM. **(b)** MAE decreasing as pressure is increased.

**Table 1.** Relative energies of different magnetic phases of MST in meV/Mn.

|  | ΔE (meV/Mn) U=0.0 eV | | | | | |
|---|---|---|---|---|---|---|
| **Pressure (GPa)** | **0** | **0.25** | **0.5** | **0.75** | **1** | **Reported [13] (0 GPa)** |
| **FI1(GS)** | 0 | 0 | 0 | 0 | 0 | 0 |
| **AF1** | 23.36 | 25.36 | 27.44 | 30.11 | 32.92 | 19.1 |
| **AF2** | 33.29 | 34.02 | 35.1 | 35.96 | 37.19 | 31.5 |
| **FI2** | 32.97 | 34.57 | 36.37 | 38.37 | 40.59 | 32.2 |
| **FM** | 102.98 | 104.3 | 105.9 | 107.1 | 108.6 | 105.4 |
|  | ΔE (meV/Mn) U=3.0 eV | | | | | |
| **FI1(GS)** | 0 | 0 | 0 | 0 | 0 | - |
| **AF1** | 15.96 | 17.37 | 18.77 | 20.63 | 22.49 | - |
| **AF2** | 13.61 | 14.42 | 15.32 | 16.10 | 17.19 | - |
| **FI2** | 16.28 | 17.41 | 18.64 | 20.0 | 21.54 | - |
| **FM** | 43.98 | 46.62 | 49.48 | 52.74 | 56.42 | 43 |

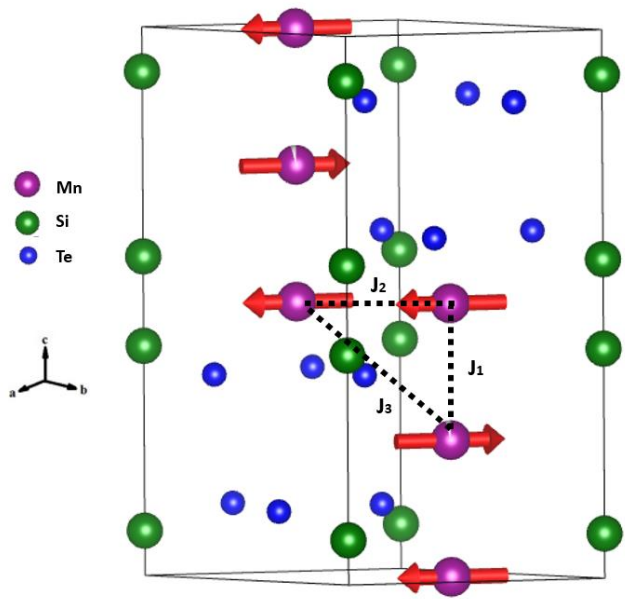

**Figure 5** The crystal structure of $Mn_3Si_2Te_6$. FI1 is the ground state whose magnetic configuration examined by first principles calculation is shown.

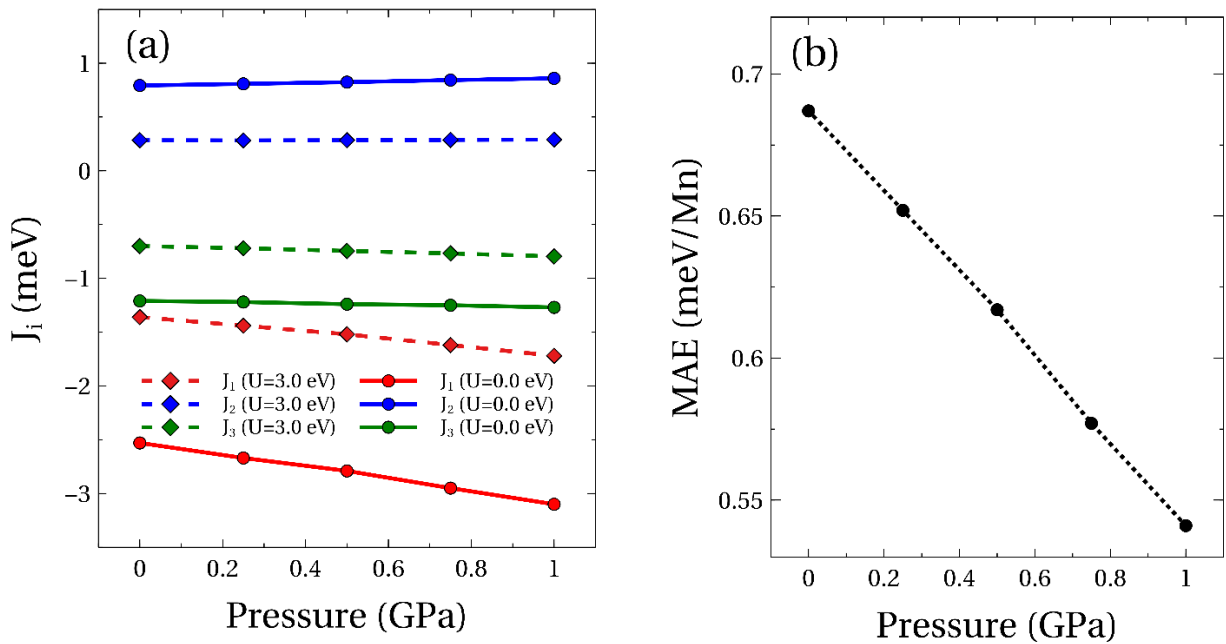

**Figure 6** Pressure dependence of the **(a)** exchange coupling parameters for MST

and **(b)** MAE.

Supplementary Information:

Pressure-induced magnetic properties of quasi-2D $Cr_2Si_2Te_6$ and $Mn_3Si_2Te_6$

Rubyann Olmos[1†], Po-Hao Chang[1], Prakash Mishra[2], Rajendra R. Zope[1], Tunna Baruah[1], Cedomir Petrovic[3], Yu Liu[3‡], Srinivasa R. Singamaneni[1*]

[1]Department of Physics, The University of Texas at El Paso, El Paso, TX 79968, USA

[2]Computational Science Program, The University of Texas at El Paso, El Paso, TX 79968, USA

[3]Condensed Matter Physics and Materials Science Department, Brookhaven National Laboratory, Upton, New York 11973, USA

*Corresponding author: srao@utep.edu

†Present address: Department of Physics and Astronomy, Rice University, Houston, TX 77005, USA.

‡Present address: Los Alamos National Laboratory, Los Alamos, NM 87545, USA

CST sample was fixed flat using carbon tape on the Teflon cap which is then inserted into the sample chamber. However, MST exhibits an easy axis of H || *ab*, therefore the sample must be fixed in a "vertical" fashion. For this case we align the sample using two pieces of carbon tape sandwiching a small portion of the sample, while still allowing for the Daphne oil to surround the sample to give a hydrostatic pressure effect. In this orientation, it is important to know the length of the sample so that clearance is known so that the sample is not crushed as the length of the cell is increasingly compressed when adding pressure.

The sample is carefully mounted in the easy axis direction with carbon tape securing the edge to the Teflon cap. The scenario in which MST is fixed is different than those materials whose easy axis is for H || *c*. Crystals with out-of-plane magnetic moments can easily be fixed flat against the Teflon cap. However, in the case of MST the sample will need be mounted vertically to have magnetic fields parallel to the *ab*-plane (H || *ab*). Therefore, it is important to note down the length of the sample so that one does not crush the crystal upon compression of the pressure cell. It also important to note that the sample may move slightly, therefore inspection of the orientation and whether the crystal remained intact should be made upon decompression of the pressure cell.

The pressure is determined by measuring the moment of Pb as a function of temperature between 6.8–7.3 K with a 1 Gauss magnetic field. The derivative of the moment with temperature reveals a minimum, indicating the magnetic transition temperature of the Pb. It is already well-known how the Pb manometer behaves with pressure following the rate of 0.379 K/GPa. As the ambient transition temperature for Pb lies between 7.185-7.19 K, calculating the change in transition temperature from the applied pressure to zero pressure will allow one to calculate the pressure inside the cell using the expression: (7.19 [K] – $T_C$ [K]) / 0.379 [K/GPa].

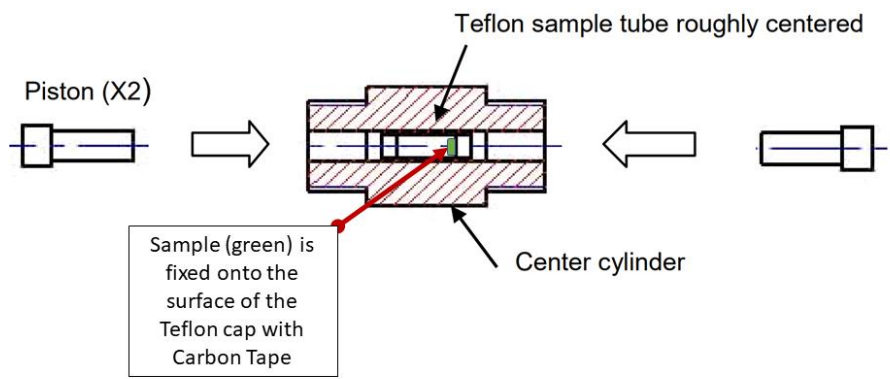

**Figure S1** *Depiction of the center cylinder and pistons of the BeCu pressure cell. Inside the center cylinder is a Teflon tube with Teflon caps on each end.*